\documentstyle[11pt]{article}
\begin{document}

\centerline{\Large \bf A puzzle in quantum dynamics}
\vskip 30pt

\centerline{C. Y. Chen}
\centerline{Dept. of Physics, Beijing University of Aeronautics}
\centerline{and Astronautics, Beijing 100083, PRC}
\vskip 20pt
\centerline{Email: cychen@public2.east.net.cn}

\vfill
\noindent {\bf Abstract:}
The textbook treatment in that the wave function of a dynamical system 
is expanded in an eigenfunction series is investigated. 
With help of an elementary example and some mathematical theorems, it is 
revealed that in terms of solving the time-dependent Schr\"odinger 
equation the treatment involves in divergence trouble . The 
root reason behind the trouble is finally analyzed.

\vskip 10pt
\noindent PACS numbers: 03.65-w
\newpage
Solutions of the time-dependent Schr\"odinger equation, formal ones and 
numerical ones, are of considerable interest in the past and in the present. 
    It is recognized that they are related to many interesting and essential 
 dynamical processes in the quantum realm.

Being aware of difficulties with solving the time-dependent Schr\"odinger 
equation directly, the textbook treatment takes the detour-type approach in 
that a wave function is expanded in an eigenfunction series and thus 
the dynamics of coefficients of the series, instead of that of the wave 
function, becomes the focus of the theory\cite{text}. The procedure was put 
forward initially by Dirac at the early time of quantum mechanics and has 
since been accepted as one of standard methods in many perturbative and 
nonperturbative approaches.

Though not enough attention was received, discussions and suggestions 
concerning the Dirac approach existed for 
decades\cite{decades}\cite{decades1}. In some of our recent papers, it was 
argued that even for today's physicists there were still things worth 
careful consideration\cite{chen0}. The purpose of this paper is to examine 
the pure mathematical aspect of the subject. By the term ``the pure 
mathematical aspect'' we mean that (i) the foundation of solving the 
time-dependent Schr\"odinger equation, not any concrete perturbative 
procedure, will be of interest; (ii) only simple examples that can be 
investigated both numerically and analytically will be 
discussed; (iii) except the energy concept no physical arguments, 
particularly those of gauge, will be related.

The basic assumption of the standard treatment is that a system's 
wavefunction can be expanded in a series 
of the following form 
\begin{equation}\label{solu}\Psi(t,{\bf r})= \sum C_n(t) e^{-iE_n 
t/\hbar} 
\Psi_n({\bf r}),\end{equation}
 where $C_n(t)$ stands for a coefficient that is a constant before the 
initial time and then becomes time-dependent after it and $\Psi_n({\bf 
r})$ is one of the eigenfunctions satisfying the eigenvalue equation
\begin{equation} \label{h0} H_0 \Psi_n({\bf r})= E_n \Psi_n({\bf r}), 
\end{equation}
in which $H_0$, according to the convention, represents the Hamiltonian of 
the system at the 
initial time. Provided that all eigenfunctions are adequately normalized, 
the coefficients of the expansion must obey the normalization condition 
\begin{equation} \label{norm} \sum\limits_n^\infty |C_n(t)|^2 =1. 
\end{equation}
Inserting (\ref{solu}) into the time-dependent 
Schr\"odinger equation
\begin{equation}\label{seq} i\hbar\frac{\partial \Psi}{\partial t} = 
H(t,{\bf r}) \Psi, \end{equation}
multiplying both the sides by $e^{iE_n t/\hbar}\Psi_n^*({\bf r})$ and 
integrating the resultant equation term by term, we arrive at a set of 
coupled ordinary-differential equations 
\begin{equation}\label{ordi} 
i\hbar \frac{dC_n}{dt} =\sum\limits_l C_l V_{nl} \exp (i\omega_{nl} t) , 
\end{equation} 
 where $\omega_{nl}=(E_n-E_l)/\hbar$ and $V_{nl}$ stands for the matrix 
element of the Hamiltonian variation $V(t) \equiv H(t)-H_0$. It is almost 
unanimously believed that the equation set (\ref{ordi}) is completely and 
exactly equivalent to the Schr\"odinger equation (\ref{seq})
and, at the same time, 
enjoys some advantages. One of the advantages is that the equation set 
appears to be numerically solvable and some efforts are indeed in this 
direction\cite{decades1}. 
According to the customary thought, there is nothing to be discussed 
concerning the validity of (\ref{ordi}) because (i) no approximation has 
been introduced; (ii) no perturbative procedure has been invoked; (iii) no 
physical interpretation has been imposed. 

Nevertheless, we will, in this paper, show that the formalism outlined 
above is not truly reliable since the basic expansion (\ref{solu}) 
is not a mathematically well-behaved series.

Before discussing the issue in an abstract way, let's apply the standard 
formalism to a one-dimensional oscillator and see where it leads us to. 
The oscillator is assumed to have the Hamiltonian 
\begin{equation}\label{ht} H(t)=\frac {p^2}{2 m} + \frac {S(t)}
2  k^2 x^2,\end{equation}
where  
\begin{equation} S(t)=\left\{ \begin{array}{ll} 1& (t\leq 0)  \\ 1+ \eta t 
&(0<t<T) \\1+\eta T &(t\geq T). \end{array} \right.\end{equation}
(Several other types of time-dependent Hamiltonians were also tried. 
The situations were all alike as far as the subject of this 
paper was concerned.) It is well-known that the system expressed by 
(\ref{ht}) is a simple harmonic oscillator before $t=0$.
 The quantum state before $t=0$ is expressed by
\begin{equation} \label{cn0}
\sum\limits_{n=0} C_{n0} e^{-i(n+1/2)\omega t} N_n(\alpha) H_n(\alpha x) 
e^{-\frac 12 \alpha^2 x^2},\end{equation}
where $\omega=\sqrt{k/m}$, $\alpha=(mk/\hbar^2)^{1/4}$, $N_n=[\alpha/
(\sqrt \pi 2^n n!)]^{1/2}$ and $H_n$ is the $n$th Hermite polynomial, 
which is governed by 
the iteration relation
\begin{equation}\label{iteration}
 H_{n+1}(\xi)= 2\xi H_n(\xi) - 2n H_{n-1}(\xi) \end{equation}
with
\begin{equation} H_0(\xi)=1,\quad H_1(\xi)=2\xi, \quad H_2(\xi)=4\xi^2-2
\end{equation}

We will assume that the system is in the ground state before $t=0$, namely 
we have 
\begin{equation}
\label{inic} C_0(0)=1\quad {\rm and}\quad  C_n(0)=0\; (n\not= 0),
\end{equation}
and try to calculate the quantum state at the later times.

To make the numerical work as simple as possible, we set
\begin{equation}\label{pa} \eta =1,\;T=1,\; m=1,\;k=1,\;\hbar=1.
\end{equation}
Under these conditions, the Hamiltonian variation $V(t)$ becomes
\begin{equation}\label{vt} V(t) = 
\left\{ \begin{array}{ll}  x^2 t/2 &(0<t<T) \vspace{4pt}  \\ 
  x^2/2 &(t\geq T).
\end{array} \right.
\end{equation}
By virtue of the iteration relation (\ref{iteration}), we obtain the 
following relations 
\begin{equation}\label{x21}
\langle 2j|x^2/2|2j \rangle= \left( j+0.25\right) 
\end{equation}
 and
\begin{equation}\label{x22} \langle 2j|x^2/2|2j+2\rangle= 
0.5\times [(j+0.5)(j+1)]^{\frac 12}. \end{equation} 
With the notation $C_n=C_n^r+iC_n^i$, the equation set (\ref{ordi}) becomes  
    \begin{equation}\label{formr}\begin{array}{l}\displaystyle
 \frac{dC_{2j}^r}{dt} =[C_{2j-2}^i\cos(2t)+C_{2j-2}^r\sin(2t)] V_{2j-2,2j}
 \vspace{4pt}\\
\qquad+C_{2j}^i V_{2j,2j}+ [C_{2j+2}^i \cos(2t)-C_{2j+2}^r\sin(2t)] V_{2j,
2j+2}\end{array}\end{equation}
and 
\begin{equation}\label{formi}\begin{array}{l}\displaystyle
 \frac{dC_{2j}^i}{dt}=-\{[C_{2j-2}^r\cos(2t)-C_{2j-2}^i\sin(2t)]
 V_{2j-2,2j}\vspace{4pt}\\
\qquad +C_{2j}^r V_{2j,2j}+
[C_{2j+2}^r\cos(2t)+C_{2j+2}^i\sin(2t)]V_{2j,2j+2}\},
\end{array}\end{equation} 
 where $j$ runs over all non-negative integers. Note that in the equations 
above any quantity having a negative subindex actually vanishes. Namely, 
$C^r_{-2}=C^i_{-2}=0$ and $V_{-2,0}=0$. 
 
In numerical mathematics the Runge-Kutta algorithm\cite{num} says that for
a vector-type differential equation 
 \begin{equation} \dot {\bf y}= {\bf f}({\bf y},t), \end{equation}
we can construct the solution according to the formula
\begin{equation}\label{rk}
{\bf y}_{j+1}={\bf y}_j+\frac 16({\bf z}_1+2
 {\bf z}_2+2 {\bf z}_3+{\bf z}_4),\end{equation}
where 
\begin{equation} {\bf y}_0={\bf y}(t_0),\;{\bf y}_1={\bf y}(t_1),\;
\cdots,{\bf y}_j={\bf y}(t_j),\cdots,\end{equation}
with $t_j-t_{j-1}=h$, and 
$$\begin{array}{l} {\bf z}_1=h{\bf f}({\bf y}_j,t_j)\\
 {\bf z}_2=h{\bf f}({\bf y}_j+{\bf z}_1/2,t_j+h/2)\\
 {\bf z}_3=h{\bf f}({\bf y}_j+{\bf z}_2/2,t_j+h/2)\\
 {\bf z}_4=h{\bf f}({\bf y}_j+{\bf z}_3,t_j+h).\end{array}$$
Although the algorithm proves very effective in classical 
mechanics\cite{chen1}, it encounters difficulties one by one in dealing 
with (\ref{formr}) and (\ref{formi}). Firstly, it is found that the number 
of involved variables, namely nonzero $C_{2j}^r$ and $C_{2j}^i$, increases 
rapidly. At $t=0$, only $C_0^r=1$ is there. At $t=nh$, we have $(8n+2)$ 
nonzero variables. Secondly, when the code, adapted to the increasing 
number of variables, is run in a computer, it is found that if the 
Hamiltonian $V(t)$ is relatively large, the normalization condition 
$\sum |C_n(t)|^2=1$ will soon be broken down. Figure 1 illustrates that 
the breakdown takes place within an oscillation cycle in respect to the 
perturbation described by (\ref{vt}) and $h=0.001$. Thirdly, it is noted 
that whenever the normalization condition 
runs into difficulty, the average energy of the system 
\begin{equation}\label{energy} 
\langle E \rangle = \sum E_n|C_n(t)|^2 \end{equation}
tends to infinity in a similar way as shown in Figure 2. (The code is 
available if more details are interested.)

If we study the numerical calculation in an analytical manner, more 
puzzling things surface impressively. Equations 
(\ref{formr}) and (\ref{formi}) suggest that if the system is in the ground 
state initially, it will, at the next step, 
be partly in the second excited state owning to the 
coupling of the coefficients. At the next next step, it will be partly 
in the fourth excited state. Since the Hamiltonian variation $V(t)$ 
 (\ref{vt}) has a finite value as $t\rightarrow \infty$, the 
process will never terminate and the state of the system will vary forever. 
However, elementary quantum 
mechanics\cite{schiff} asserts that the system will be settled in a 
stationary state after $t=T$. The contrast becomes more striking if we look 
at Figure 2, in which the ``energy'' curve rises and falls at the times 
$t>T=1$ disregarding any theory about stationary states.

Another perplexing thing is related to energy more directly.
According to the formulas (\ref{norm}) and (\ref{energy}), 
a ground-state system cannot lower its energy under any circumstances. 
The physical intuition, however, tells us a different story. 
For instance, if the Hamiltonian variation of our oscillator takes 
the following form 
\begin{equation}\label{vt1} V(t) = 
- x^2 t/2 \quad (t>0),\end{equation}
the system, in contrast 
to the system affected by (\ref{vt}), should lose its energy since, 
roughly speaking, the perturbation (\ref{vt1}) makes the potential 
energy of the system smaller while leaving the kinetic energy unchanged. 

We now try to work out the puzzle related to the title of this paper: Why
does the standard formalism, derived from the Schr\"odinger equation in a 
seemingly rigorous way, suffer from so many difficult things? For our 
purposes, the analysis will 
be done almost entirely from mathematical consideration.

It should be pointed out that  
\begin{equation}
e^{-iE_0t/\hbar}\Psi_0({\bf r}),e^{-iE_1t/\hbar}\Psi_1({\bf r}),
\cdots,e^{-iE_nt/\hbar}\Psi_n({\bf r}),\cdots
\end{equation}
do not constitute an orthogonal-and-complete basis for a function 
$f(t,{\bf r})$. Consider our one-dimensional case again. Many 
functions, such as
\begin{equation} t,\; (t+x),\; \cos(t x)e^{-\frac 12 \alpha^2 x^2 },\;
 e^{tx}e^{-\frac 12 \alpha^2 x^2 }, \;\cdots\cdots,
\end{equation}
cannot be expanded in a series of the form (\ref{cn0}).
That is to say, the practice of expanding the solution of a Schr\"odinger 
equation in a series of eigenfunctions is originated from an intuitive 
physical idea rather than from a stringent mathematical analysis. 

The major argument, which we wish to propose in this paper, is the 
following. Even if the wave function at any fixed time can be expanded in an 
 eigenfunction series of the form   
\begin{equation}\label{solu1} \sum C_n(t) \Psi_n({\bf r}), 
\end{equation}
substitution of this expansion, or the equivalent series (\ref{solu}), into 
the time-dependent Schr\"odinger equation is still a questionable procedure.
As a matter of fact, there indeed exist mathematical theorems restricting 
the use of infinite series\cite{rudin}. One of them states that only a 
series converging uniformly can safely be integrated term by term. Another 
of them indicates that to differentiate a series term by term the 
resultant series has to be uniformly convergent. An inspection can tell 
us that any known eigenfunction series, if given as a truly infinite one, 
behaves poorly in this respect and cannot be safely used in the 
time-dependent Schr\"odinger equation. To see this in an example,
apply the standard treatment to the series (\ref{cn0}). 
Assume that in the series
\begin{equation}\label{cn00}
C_{n0}=\frac{\sqrt 6}{\pi(n+1)},\quad{\rm such\; that}\quad \sum 
|C_{n0}|^2=\frac 6{\pi^2} \sum \frac 1{(n+1)^2}=1. \end{equation}
By taking the time derivative, the left side of the Schr\"odinger 
equation becomes 
\begin{equation} \label{cn01}
i\hbar \frac{\partial \Psi}{\partial t}= \frac{\sqrt{6} \hbar\omega}
{\pi} e^{-\frac 12 \alpha^2 x^2}\left[\sum_n^\infty\frac {n+1/2}{n+1}
e^{-i(n+1/2)\omega t} N_n(\alpha)H_n(\alpha x)\right] .\end{equation}
For this step and the next step to make sense, in which the expression 
has been differentiated and will be integrated term by term, 
it is necessary to
ensure the uniform convergence of the series included in the square 
bracket above. (Note that the common factor $e^{-(\alpha x)^2/2}$
is not relevant to the treatment term by term.)
By setting $t=0$, this series becomes 
\begin{equation}\label{seriesf}
\sum_n^\infty \frac{n+1/2}{n+1} N_n(\alpha)H_n(\alpha x)\sim 
\sum_n^\infty \frac{(2\alpha x)^n}{(2^nn!)^{1/2}}, \end{equation} 
whose value rises very fast for large $x$ (faster than 
$e^{2\alpha x}$). A simple analysis tells us that the series cannot
be uniformly convergent since it has no upper limit on its domain.

At this stage, it is desired to have explanation for the
fact that the series form 
\begin{equation}\label{sexp}
\sum C_{n0} e^{-iE_n t/\hbar} \Psi_n({\bf r}), 
\end{equation}
where $C_{n0}$ are pure time-independent constants, has safely been used
for almost all stationary states. We believe that there are two facts 
responsible for it. The first one is that for practical cases the initial 
expansion of (\ref{sexp}) takes the form   
\begin{equation}\label{ini} \sum_{n=0}^N C_{n0} \Psi_n({\bf r}) 
,\end{equation}
 where $N$ is a finite number. The finiteness of the series is due to the 
fact that high-energy eigenfunctions are related to the wave function in 
the very remote regions, which are usually of no interest for the problem, 
in which a fixed accuracy is set up at the very beginning. The second one
is that each term in (\ref{ini}) evolves in the Schr\"odinger equation 
independently (no coupling), and therefore the series, if finite 
initially, keeps finite in all treatments. Keeping these facts in mind, 
one may say that a stationary wave function can indeed be expanded in 
a series of eigenfunctions (with limitations still, see Ref. 4).

The discussion above has almost already manifested that things will be
different for nonstationary cases. We may still express the dynamical 
wave function as a finite series at the initial time, but the series soon 
becomes a truly infinite one. The strong coupling, as (\ref{ordi}) 
exhibits, tends to make each of the coefficients have a significant value.
This is disastrous in view of that any truly infinite eigenfunction series 
suffers seriously in the time-dependent Schr\"odinger equation, as has 
been unveiled.

From a slightly different viewpoint, we can also see that the expansion
(\ref{solu}) cannot work smoothly. It is obvious that after the initial 
time the eigenfunctions defined by $H_0$ are no longer 
adequate eigenfunctions and the eigenfrequencies defined by 
$\omega_n=E_n/\hbar$ are no longer adequate eigenfrequencies. If we 
forcefully express the dynamical wave function in a series constructed 
from these ``out-of-date'' eigenfunctions and ``out-of-date'' 
eigenfrequencies, the coefficients of the series have to adjust 
themselves violently. Since the large scale (the entire space)  
and the fast variation (represented by $e^{-in\omega t}$) get involved, 
the adjustment has to be so violent that the normalization condition 
$\sum |C_n|^2=1$ will definitely break down. 

We proposed two alternatives to the standard 
treatment\cite{chen2}\cite{chen3}. 
One is based on a perturbative approach and the 
other is based on a nonperturbative approach. In the nonperturbative
approach, with help of ``intermediate'' eigenfunctions and  
eigenfrequencies a general procedure of solving the time-dependent 
Schr\"odinger equation is constructed. 

Mathematical discussion with Professor Qihou Liu is gratefully 
acknowledged.   
The work is partly supported by the fund provided by Education 
Ministry, P.R. China.

\newpage
\centerline{\LARGE \bf  Figure captions}
\vskip 10pt
\noindent Fig. 1, The time behavior of the coefficients in terms of the 
normalization condition. Note that the period of the oscillator is 6.28 
since $\omega=1$.
\vskip 5pt
\noindent Fig. 2, The time behavior of the coefficients in terms of the 
average energy.

\newpage
\noindent {\bf Figure 1}          

\setlength{\unitlength}{0.018in} 
\begin{picture}(250,160)

\put(86,11){\makebox(8,8)[c]{\small 1}}
\put(126,11){\makebox(8,8)[c]{\small 2}}
\put(166,11){\makebox(8,8)[c]{\small 3}}
\put(186,16){\makebox(8,8)[c]{$t$}}
\multiput(90,20)(40,0){3}{\line(0,1){5}}
\multiput(70,20)(40,0){3}{\line(0,1){3}}

\put(37,61){\makebox(8,8)[c]{\small 1.0}}
\put(37,101){\makebox(8,8)[c]{\small 100}}
\put(30,118){\makebox(40,20)[c]{\small $\sum |C_n|^2$}}

\put(50,105){\line(1,0){4}}
\put(50,15){\line(0,1){65}} 
\put(50,90){\line(0,1){29}} 
\put(50.2,119){\vector(0,1){1}} 
\multiput(50,82.5)(0,2.5){3}{\circle*{1}}

\put(45,20){\line(1,0){139}}
\put(183,20.4){\vector(1,0){2}}

\put(50.000, 65.000){\circle*{1}}
\put(50.400, 65.000){\circle*{1}}
\put(50.800, 65.000){\circle*{1}}
\put(51.200, 65.000){\circle*{1}}
\put(51.600, 65.000){\circle*{1}}
\put(52.000, 65.000){\circle*{1}}
\put(52.400, 65.000){\circle*{1}}
\put(52.800, 65.000){\circle*{1}}
\put(53.200, 65.000){\circle*{1}}
\put(53.600, 65.000){\circle*{1}}
\put(54.000, 65.000){\circle*{1}}
\put(54.400, 65.000){\circle*{1}}
\put(54.800, 65.000){\circle*{1}}
\put(55.200, 65.000){\circle*{1}}
\put(55.600, 65.000){\circle*{1}}
\put(56.000, 65.000){\circle*{1}}
\put(56.400, 65.000){\circle*{1}}
\put(56.800, 65.000){\circle*{1}}
\put(57.200, 65.000){\circle*{1}}
\put(57.600, 65.000){\circle*{1}}
\put(58.000, 65.000){\circle*{1}}
\put(58.400, 65.000){\circle*{1}}
\put(58.800, 65.000){\circle*{1}}
\put(59.200, 65.000){\circle*{1}}
\put(59.600, 65.000){\circle*{1}}
\put(60.000, 65.000){\circle*{1}}
\put(60.400, 65.000){\circle*{1}}
\put(60.800, 65.000){\circle*{1}}
\put(61.200, 65.000){\circle*{1}}
\put(61.600, 65.000){\circle*{1}}
\put(62.000, 65.000){\circle*{1}}
\put(62.400, 65.000){\circle*{1}}
\put(62.800, 65.000){\circle*{1}}
\put(63.200, 65.000){\circle*{1}}
\put(63.600, 65.000){\circle*{1}}
\put(64.000, 65.000){\circle*{1}}
\put(64.400, 65.000){\circle*{1}}
\put(64.800, 65.000){\circle*{1}}
\put(65.200, 65.000){\circle*{1}}
\put(65.600, 65.000){\circle*{1}}
\put(66.000, 65.000){\circle*{1}}
\put(66.400, 65.000){\circle*{1}}
\put(66.800, 65.000){\circle*{1}}
\put(67.200, 65.000){\circle*{1}}
\put(67.600, 65.000){\circle*{1}}
\put(68.000, 65.000){\circle*{1}}
\put(68.400, 65.000){\circle*{1}}
\put(68.800, 65.000){\circle*{1}}
\put(69.200, 65.000){\circle*{1}}
\put(69.600, 65.000){\circle*{1}}
\put(70.000, 65.000){\circle*{1}}
\put(70.400, 65.000){\circle*{1}}
\put(70.800, 65.000){\circle*{1}}
\put(71.200, 65.000){\circle*{1}}
\put(71.600, 65.000){\circle*{1}}
\put(72.000, 65.000){\circle*{1}}
\put(72.400, 65.000){\circle*{1}}
\put(72.800, 65.000){\circle*{1}}
\put(73.200, 65.000){\circle*{1}}
\put(73.600, 65.000){\circle*{1}}
\put(74.000, 65.000){\circle*{1}}
\put(74.400, 65.000){\circle*{1}}
\put(74.800, 65.000){\circle*{1}}
\put(75.200, 65.000){\circle*{1}}
\put(75.600, 65.000){\circle*{1}}
\put(76.000, 65.000){\circle*{1}}
\put(76.400, 65.000){\circle*{1}}
\put(76.800, 65.000){\circle*{1}}
\put(77.200, 65.000){\circle*{1}}
\put(77.600, 65.000){\circle*{1}}
\put(78.000, 65.000){\circle*{1}}
\put(78.400, 65.000){\circle*{1}}
\put(78.800, 65.000){\circle*{1}}
\put(79.200, 65.000){\circle*{1}}
\put(79.600, 65.000){\circle*{1}}
\put(80.000, 65.000){\circle*{1}}
\put(80.400, 65.000){\circle*{1}}
\put(80.800, 65.000){\circle*{1}}
\put(81.200, 65.000){\circle*{1}}
\put(81.600, 65.000){\circle*{1}}
\put(82.000, 65.000){\circle*{1}}
\put(82.400, 65.000){\circle*{1}}
\put(82.800, 65.000){\circle*{1}}
\put(83.200, 65.000){\circle*{1}}
\put(83.600, 65.000){\circle*{1}}
\put(84.000, 65.000){\circle*{1}}
\put(84.400, 65.000){\circle*{1}}
\put(84.800, 65.000){\circle*{1}}
\put(85.200, 65.000){\circle*{1}}
\put(85.600, 65.000){\circle*{1}}
\put(86.000, 65.000){\circle*{1}}
\put(86.400, 65.000){\circle*{1}}
\put(86.800, 65.000){\circle*{1}}
\put(87.200, 65.000){\circle*{1}}
\put(87.600, 65.000){\circle*{1}}
\put(88.000, 65.000){\circle*{1}}
\put(88.400, 65.000){\circle*{1}}
\put(88.800, 65.000){\circle*{1}}
\put(89.200, 65.000){\circle*{1}}
\put(89.600, 65.000){\circle*{1}}
\put(90.000, 65.000){\circle*{1}}
\put(90.400, 65.000){\circle*{1}}
\put(90.800, 65.000){\circle*{1}}
\put(91.200, 65.000){\circle*{1}}
\put(91.600, 65.000){\circle*{1}}
\put(92.000, 65.000){\circle*{1}}
\put(92.400, 65.000){\circle*{1}}
\put(92.800, 65.000){\circle*{1}}
\put(93.200, 65.000){\circle*{1}}
\put(93.600, 65.000){\circle*{1}}
\put(94.000, 65.000){\circle*{1}}
\put(94.400, 65.000){\circle*{1}}
\put(94.800, 65.000){\circle*{1}}
\put(95.200, 65.000){\circle*{1}}
\put(95.600, 65.000){\circle*{1}}
\put(96.000, 65.000){\circle*{1}}
\put(96.400, 65.000){\circle*{1}}
\put(96.800, 65.000){\circle*{1}}
\put(97.200, 65.000){\circle*{1}}
\put(97.600, 65.000){\circle*{1}}
\put(98.000, 65.000){\circle*{1}}
\put(98.400, 65.000){\circle*{1}}
\put(98.800, 65.000){\circle*{1}}
\put(99.200, 65.000){\circle*{1}}
\put(99.600, 65.000){\circle*{1}}
\put(100.000, 65.000){\circle*{1}}
\put(100.400, 65.000){\circle*{1}}
\put(100.800, 65.000){\circle*{1}}
\put(101.200, 65.000){\circle*{1}}
\put(101.600, 65.000){\circle*{1}}
\put(102.000, 65.000){\circle*{1}}
\put(102.400, 65.000){\circle*{1}}
\put(102.800, 65.000){\circle*{1}}
\put(103.200, 65.000){\circle*{1}}
\put(103.600, 65.000){\circle*{1}}
\put(104.000, 65.000){\circle*{1}}
\put(104.400, 65.000){\circle*{1}}
\put(104.800, 65.000){\circle*{1}}
\put(105.200, 65.000){\circle*{1}}
\put(105.600, 65.000){\circle*{1}}
\put(106.000, 65.000){\circle*{1}}
\put(106.400, 65.000){\circle*{1}}
\put(106.800, 65.000){\circle*{1}}
\put(107.200, 65.000){\circle*{1}}
\put(107.600, 65.000){\circle*{1}}
\put(108.000, 65.000){\circle*{1}}
\put(108.400, 65.000){\circle*{1}}
\put(108.800, 65.000){\circle*{1}}
\put(109.200, 65.000){\circle*{1}}
\put(109.600, 65.000){\circle*{1}}
\put(110.000, 65.000){\circle*{1}}
\put(110.400, 65.000){\circle*{1}}
\put(110.800, 65.000){\circle*{1}}
\put(111.200, 65.000){\circle*{1}}
\put(111.600, 65.000){\circle*{1}}
\put(112.000, 65.000){\circle*{1}}
\put(112.400, 65.000){\circle*{1}}
\put(112.800, 65.000){\circle*{1}}
\put(113.200, 65.000){\circle*{1}}
\put(113.600, 65.000){\circle*{1}}
\put(114.000, 65.000){\circle*{1}}
\put(114.400, 65.000){\circle*{1}}
\put(114.800, 65.000){\circle*{1}}
\put(115.200, 65.000){\circle*{1}}
\put(115.600, 65.000){\circle*{1}}
\put(116.000, 65.000){\circle*{1}}
\put(116.400, 65.000){\circle*{1}}
\put(116.800, 65.000){\circle*{1}}
\put(117.200, 65.000){\circle*{1}}
\put(117.600, 65.000){\circle*{1}}
\put(118.000, 65.000){\circle*{1}}
\put(118.400, 65.000){\circle*{1}}
\put(118.800, 65.000){\circle*{1}}
\put(119.200, 65.000){\circle*{1}}
\put(119.600, 65.000){\circle*{1}}
\put(120.000, 65.000){\circle*{1}}
\put(120.400, 65.000){\circle*{1}}
\put(120.800, 65.000){\circle*{1}}
\put(121.200, 65.000){\circle*{1}}
\put(121.600, 65.000){\circle*{1}}
\put(122.000, 65.000){\circle*{1}}
\put(122.400, 65.000){\circle*{1}}
\put(122.800, 65.000){\circle*{1}}
\put(123.200, 65.000){\circle*{1}}
\put(123.600, 65.000){\circle*{1}}
\put(124.000, 65.000){\circle*{1}}
\put(124.400, 65.000){\circle*{1}}
\put(124.800, 65.000){\circle*{1}}
\put(125.200, 65.000){\circle*{1}}
\put(125.600, 65.000){\circle*{1}}
\put(126.000, 65.000){\circle*{1}}
\put(126.400, 65.000){\circle*{1}}
\put(126.800, 65.000){\circle*{1}}
\put(127.200, 65.000){\circle*{1}}
\put(127.600, 65.000){\circle*{1}}
\put(128.000, 65.000){\circle*{1}}
\put(128.400, 65.000){\circle*{1}}
\put(128.800, 65.000){\circle*{1}}
\put(129.200, 65.000){\circle*{1}}
\put(129.600, 65.000){\circle*{1}}
\put(130.000, 65.000){\circle*{1}}
\put(130.400, 65.000){\circle*{1}}
\put(130.800, 65.000){\circle*{1}}
\put(131.200, 65.000){\circle*{1}}
\put(131.600, 65.000){\circle*{1}}
\put(132.000, 65.000){\circle*{1}}
\put(132.400, 65.000){\circle*{1}}
\put(132.800, 65.000){\circle*{1}}
\put(133.200, 65.000){\circle*{1}}
\put(133.600, 65.000){\circle*{1}}
\put(134.000, 65.000){\circle*{1}}
\put(134.400, 65.000){\circle*{1}}
\put(134.800, 65.000){\circle*{1}}
\put(135.200, 65.000){\circle*{1}}
\put(135.600, 65.000){\circle*{1}}
\put(136.000, 65.000){\circle*{1}}
\put(136.400, 65.000){\circle*{1}}
\put(136.800, 65.000){\circle*{1}}
\put(137.200, 65.000){\circle*{1}}
\put(137.600, 65.000){\circle*{1}}
\put(138.000, 65.000){\circle*{1}}
\put(138.400, 65.000){\circle*{1}}
\put(138.800, 65.000){\circle*{1}}
\put(139.200, 65.000){\circle*{1}}
\put(139.600, 65.000){\circle*{1}}
\put(140.000, 65.000){\circle*{1}}
\put(140.400, 65.000){\circle*{1}}
\put(140.800, 65.000){\circle*{1}}
\put(141.200, 65.000){\circle*{1}}
\put(141.600, 65.000){\circle*{1}}
\put(142.000, 65.000){\circle*{1}}
\put(142.400, 65.000){\circle*{1}}
\put(142.800, 65.000){\circle*{1}}
\put(143.200, 65.000){\circle*{1}}
\put(143.600, 65.000){\circle*{1}}
\put(144.000, 65.000){\circle*{1}}
\put(144.400, 65.000){\circle*{1}}
\put(144.800, 65.000){\circle*{1}}
\put(145.200, 65.000){\circle*{1}}
\put(145.600, 65.000){\circle*{1}}
\put(146.000, 65.000){\circle*{1}}
\put(146.400, 65.000){\circle*{1}}
\put(146.800, 65.000){\circle*{1}}
\put(147.200, 65.000){\circle*{1}}
\put(147.600, 65.000){\circle*{1}}
\put(148.000, 65.000){\circle*{1}}
\put(148.400, 65.000){\circle*{1}}
\put(148.800, 65.000){\circle*{1}}
\put(149.200, 65.000){\circle*{1}}
\put(149.600, 65.000){\circle*{1}}
\put(150.000, 65.000){\circle*{1}}
\put(150.400, 65.000){\circle*{1}}
\put(150.800, 65.000){\circle*{1}}
\put(151.200, 65.000){\circle*{1}}
\put(151.600, 65.000){\circle*{1}}
\put(152.000, 65.000){\circle*{1}}
\put(152.400, 65.000){\circle*{1}}
\put(152.800, 65.000){\circle*{1}}
\put(153.200, 65.000){\circle*{1}}
\put(153.600, 65.000){\circle*{1}}
\put(154.000, 65.000){\circle*{1}}
\put(154.400, 65.000){\circle*{1}}
\put(154.800, 65.000){\circle*{1}}
\put(155.200, 65.000){\circle*{1}}
\put(155.600, 65.000){\circle*{1}}
\put(156.000, 65.000){\circle*{1}}
\put(156.400, 65.000){\circle*{1}}
\put(156.800, 65.000){\circle*{1}}
\put(157.200, 65.000){\circle*{1}}
\put(157.600, 65.000){\circle*{1}}
\put(158.000, 65.002){\circle*{1}}
\put(158.400, 65.004){\circle*{1}}
\put(158.800, 65.006){\circle*{1}}
\put(159.197, 65.041){\circle*{1}}
\put(159.595, 65.077){\circle*{1}}
\put(159.863, 65.319){\circle*{1}}
\put(159.941, 65.709){\circle*{1}}
\put(160.018, 66.099){\circle*{1}}
\put(160.095, 66.489){\circle*{1}}
\put(160.173, 66.879){\circle*{1}}
\put(160.191, 67.278){\circle*{1}}
\put(160.209, 67.678){\circle*{1}}
\put(160.227, 68.078){\circle*{1}}
\put(160.245, 68.477){\circle*{1}}
\put(160.263, 68.877){\circle*{1}}
\put(160.282, 69.276){\circle*{1}}
\put(160.300, 69.676){\circle*{1}}
\put(160.318, 70.075){\circle*{1}}
\put(160.336, 70.475){\circle*{1}}
\put(160.354, 70.875){\circle*{1}}
\put(160.372, 71.274){\circle*{1}}
\put(160.390, 71.674){\circle*{1}}
\put(160.408, 72.073){\circle*{1}}
\put(160.427, 72.473){\circle*{1}}
\put(160.445, 72.872){\circle*{1}}
\put(160.451, 73.272){\circle*{1}}
\put(160.457, 73.672){\circle*{1}}
\put(160.463, 74.072){\circle*{1}}
\put(160.469, 74.472){\circle*{1}}
\put(160.475, 74.872){\circle*{1}}
\put(160.481, 75.272){\circle*{1}}
\put(160.487, 75.672){\circle*{1}}
\put(160.493, 76.072){\circle*{1}}
\put(160.500, 76.472){\circle*{1}}
\put(160.506, 76.872){\circle*{1}}
\put(160.506, 77.272){\circle*{1}}
\put(160.507, 77.672){\circle*{1}}
\put(160.508, 78.072){\circle*{1}}
\put(160.508, 78.472){\circle*{1}}
\put(160.509, 78.872){\circle*{1}}
\put(160.510, 79.272){\circle*{1}}
\put(160.510, 79.672){\circle*{1}}
\put(160.511, 80.072){\circle*{1}}
\put(160.512, 80.472){\circle*{1}}
\put(160.513, 80.872){\circle*{1}}
\put(160.513, 81.272){\circle*{1}}
\put(160.514, 81.672){\circle*{1}}
\put(160.515, 82.072){\circle*{1}}
\put(160.515, 82.472){\circle*{1}}
\put(160.516, 82.872){\circle*{1}}
\put(160.517, 83.272){\circle*{1}}
\put(160.517, 83.672){\circle*{1}}
\put(160.518, 84.072){\circle*{1}}
\put(160.519, 84.472){\circle*{1}}
\put(160.519, 84.872){\circle*{1}}
\put(160.520, 85.272){\circle*{1}}
\put(160.521, 85.672){\circle*{1}}
\put(160.522, 86.072){\circle*{1}}
\put(160.522, 86.472){\circle*{1}}
\put(160.523, 86.872){\circle*{1}}
\put(160.524, 87.272){\circle*{1}}
\put(160.524, 87.672){\circle*{1}}
\put(160.525, 88.072){\circle*{1}}
\put(160.526, 88.472){\circle*{1}}
\put(160.526, 88.872){\circle*{1}}
\put(160.527, 89.272){\circle*{1}}
\put(160.528, 89.672){\circle*{1}}
\put(160.528, 90.072){\circle*{1}}
\put(160.529, 90.472){\circle*{1}}
\put(160.530, 90.872){\circle*{1}}
\put(160.531, 91.272){\circle*{1}}
\put(160.531, 91.672){\circle*{1}}
\put(160.532, 92.072){\circle*{1}}
\put(160.533, 92.472){\circle*{1}}
\put(160.533, 92.872){\circle*{1}}
\put(160.534, 93.272){\circle*{1}}
\put(160.535, 93.672){\circle*{1}}
\put(160.535, 94.072){\circle*{1}}
\put(160.536, 94.472){\circle*{1}}
\put(160.537, 94.872){\circle*{1}}

\end{picture}

\vskip 10pt
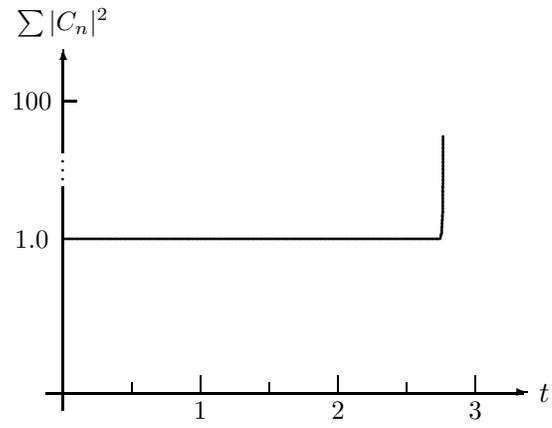
\noindent {\bf Figure 2}          

\setlength{\unitlength}{0.018in} 
\begin{picture}(250,160)

\put(86,11){\makebox(8,8)[c]{\small 1}}
\put(126,11){\makebox(8,8)[c]{\small 2}}
\put(166,11){\makebox(8,8)[c]{\small 3}}
\put(186,16){\makebox(8,8)[c]{$t$}}
\multiput(90,20)(40,0){3}{\line(0,1){5}}
\multiput(70,20)(40,0){3}{\line(0,1){3}}
\multiput(50,65)(0,20){3}{\line(1,0){4}}
\put(36,61){\makebox(8,8)[c]{\small 0.50}}
\put(36,101){\makebox(8,8)[c]{\small 0.54}}
\put(30,118){\makebox(40,20)[c]{\small $\sum E_n|C_n|^2$}}

\put(50,15){\line(0,1){20}} 
\put(50,50){\line(0,1){69}}
\put(50.2,119){\vector(0,1){1}} 

\multiput(50,39)(0,4){3}{\circle*{0.5}}

\put(45,20){\line(1,0){139}}
\put(183,20.4){\vector(1,0){2}}

\put(50.000, 65.000){\circle*{1}}
\put(50.400, 65.000){\circle*{1}}
\put(50.800, 65.000){\circle*{1}}
\put(51.200, 65.000){\circle*{1}}
\put(51.600, 65.000){\circle*{1}}
\put(52.000, 65.001){\circle*{1}}
\put(52.400, 65.001){\circle*{1}}
\put(52.800, 65.001){\circle*{1}}
\put(53.200, 65.002){\circle*{1}}
\put(53.600, 65.003){\circle*{1}}
\put(54.000, 65.004){\circle*{1}}
\put(54.400, 65.006){\circle*{1}}
\put(54.800, 65.008){\circle*{1}}
\put(55.200, 65.011){\circle*{1}}
\put(55.600, 65.015){\circle*{1}}
\put(56.000, 65.018){\circle*{1}}
\put(56.400, 65.024){\circle*{1}}
\put(56.800, 65.029){\circle*{1}}
\put(57.200, 65.037){\circle*{1}}
\put(57.600, 65.045){\circle*{1}}
\put(58.000, 65.053){\circle*{1}}
\put(58.399, 65.066){\circle*{1}}
\put(58.799, 65.078){\circle*{1}}
\put(59.199, 65.094){\circle*{1}}
\put(59.599, 65.111){\circle*{1}}
\put(59.998, 65.127){\circle*{1}}
\put(60.398, 65.150){\circle*{1}}
\put(60.797, 65.172){\circle*{1}}
\put(61.196, 65.200){\circle*{1}}
\put(61.595, 65.229){\circle*{1}}
\put(61.994, 65.257){\circle*{1}}
\put(62.392, 65.294){\circle*{1}}
\put(62.790, 65.331){\circle*{1}}
\put(63.188, 65.376){\circle*{1}}
\put(63.585, 65.422){\circle*{1}}
\put(63.983, 65.467){\circle*{1}}
\put(64.379, 65.523){\circle*{1}}
\put(64.775, 65.579){\circle*{1}}
\put(65.169, 65.646){\circle*{1}}
\put(65.564, 65.712){\circle*{1}}
\put(65.958, 65.779){\circle*{1}}
\put(66.350, 65.859){\circle*{1}}
\put(66.742, 65.938){\circle*{1}}
\put(67.131, 66.031){\circle*{1}}
\put(67.520, 66.123){\circle*{1}}
\put(67.909, 66.215){\circle*{1}}
\put(68.295, 66.322){\circle*{1}}
\put(68.680, 66.430){\circle*{1}}
\put(69.061, 66.551){\circle*{1}}
\put(69.442, 66.673){\circle*{1}}
\put(69.823, 66.794){\circle*{1}}
\put(70.199, 66.932){\circle*{1}}
\put(70.574, 67.070){\circle*{1}}
\put(70.950, 67.207){\circle*{1}}
\put(71.318, 67.363){\circle*{1}}
\put(71.686, 67.519){\circle*{1}}
\put(72.048, 67.689){\circle*{1}}
\put(72.410, 67.860){\circle*{1}}
\put(72.772, 68.030){\circle*{1}}
\put(73.126, 68.217){\circle*{1}}
\put(73.479, 68.404){\circle*{1}}
\put(73.833, 68.590){\circle*{1}}
\put(74.177, 68.793){\circle*{1}}
\put(74.522, 68.997){\circle*{1}}
\put(74.866, 69.200){\circle*{1}}
\put(75.201, 69.419){\circle*{1}}
\put(75.535, 69.639){\circle*{1}}
\put(75.869, 69.858){\circle*{1}}
\put(76.193, 70.093){\circle*{1}}
\put(76.517, 70.328){\circle*{1}}
\put(76.841, 70.562){\circle*{1}}
\put(77.155, 70.810){\circle*{1}}
\put(77.468, 71.059){\circle*{1}}
\put(77.782, 71.307){\circle*{1}}
\put(78.085, 71.568){\circle*{1}}
\put(78.388, 71.829){\circle*{1}}
\put(78.692, 72.089){\circle*{1}}
\put(78.985, 72.361){\circle*{1}}
\put(79.278, 72.633){\circle*{1}}
\put(79.571, 72.905){\circle*{1}}
\put(79.864, 73.177){\circle*{1}}
\put(80.146, 73.461){\circle*{1}}
\put(80.428, 73.745){\circle*{1}}
\put(80.710, 74.029){\circle*{1}}
\put(80.982, 74.322){\circle*{1}}
\put(81.254, 74.615){\circle*{1}}
\put(81.526, 74.908){\circle*{1}}
\put(81.798, 75.202){\circle*{1}}
\put(82.060, 75.504){\circle*{1}}
\put(82.322, 75.806){\circle*{1}}
\put(82.584, 76.109){\circle*{1}}
\put(82.846, 76.411){\circle*{1}}
\put(83.098, 76.721){\circle*{1}}
\put(83.350, 77.032){\circle*{1}}
\put(83.602, 77.343){\circle*{1}}
\put(83.854, 77.653){\circle*{1}}
\put(84.097, 77.971){\circle*{1}}
\put(84.340, 78.289){\circle*{1}}
\put(84.583, 78.606){\circle*{1}}
\put(84.826, 78.924){\circle*{1}}
\put(85.060, 79.248){\circle*{1}}
\put(85.295, 79.572){\circle*{1}}
\put(85.529, 79.896){\circle*{1}}
\put(85.764, 80.220){\circle*{1}}
\put(85.991, 80.549){\circle*{1}}
\put(86.218, 80.879){\circle*{1}}
\put(86.445, 81.208){\circle*{1}}
\put(86.672, 81.537){\circle*{1}}
\put(86.899, 81.867){\circle*{1}}
\put(87.119, 82.201){\circle*{1}}
\put(87.338, 82.535){\circle*{1}}
\put(87.558, 82.870){\circle*{1}}
\put(87.777, 83.204){\circle*{1}}
\put(87.991, 83.542){\circle*{1}}
\put(88.204, 83.881){\circle*{1}}
\put(88.417, 84.219){\circle*{1}}
\put(88.631, 84.557){\circle*{1}}
\put(88.844, 84.896){\circle*{1}}
\put(89.052, 85.238){\circle*{1}}
\put(89.260, 85.579){\circle*{1}}
\put(89.467, 85.921){\circle*{1}}
\put(89.675, 86.263){\circle*{1}}
\put(89.883, 86.605){\circle*{1}}
\put(90.087, 86.949){\circle*{1}}
\put(90.292, 87.292){\circle*{1}}
\put(90.496, 87.636){\circle*{1}}
\put(90.701, 87.980){\circle*{1}}
\put(90.905, 88.324){\circle*{1}}
\put(91.108, 88.668){\circle*{1}}
\put(91.311, 89.013){\circle*{1}}
\put(91.515, 89.357){\circle*{1}}
\put(91.718, 89.702){\circle*{1}}
\put(91.921, 90.047){\circle*{1}}
\put(92.124, 90.391){\circle*{1}}
\put(92.327, 90.736){\circle*{1}}
\put(92.530, 91.081){\circle*{1}}
\put(92.733, 91.425){\circle*{1}}
\put(92.936, 91.770){\circle*{1}}
\put(93.140, 92.114){\circle*{1}}
\put(93.344, 92.458){\circle*{1}}
\put(93.548, 92.802){\circle*{1}}
\put(93.751, 93.147){\circle*{1}}
\put(93.955, 93.491){\circle*{1}}
\put(94.161, 93.834){\circle*{1}}
\put(94.366, 94.177){\circle*{1}}
\put(94.572, 94.520){\circle*{1}}
\put(94.777, 94.863){\circle*{1}}
\put(94.983, 95.207){\circle*{1}}
\put(95.190, 95.548){\circle*{1}}
\put(95.398, 95.890){\circle*{1}}
\put(95.606, 96.232){\circle*{1}}
\put(95.814, 96.574){\circle*{1}}
\put(96.025, 96.913){\circle*{1}}
\put(96.236, 97.253){\circle*{1}}
\put(96.447, 97.593){\circle*{1}}
\put(96.658, 97.933){\circle*{1}}
\put(96.868, 98.273){\circle*{1}}
\put(97.083, 98.610){\circle*{1}}
\put(97.298, 98.948){\circle*{1}}
\put(97.513, 99.285){\circle*{1}}
\put(97.728, 99.622){\circle*{1}}
\put(97.943, 99.960){\circle*{1}}
\put(98.163, 100.294){\circle*{1}}
\put(98.383, 100.628){\circle*{1}}
\put(98.603, 100.962){\circle*{1}}
\put(98.824, 101.296){\circle*{1}}
\put(99.049, 101.626){\circle*{1}}
\put(99.275, 101.956){\circle*{1}}
\put(99.501, 102.286){\circle*{1}}
\put(99.727, 102.616){\circle*{1}}
\put(99.953, 102.947){\circle*{1}}
\put(100.186, 103.272){\circle*{1}}
\put(100.419, 103.597){\circle*{1}}
\put(100.652, 103.922){\circle*{1}}
\put(100.885, 104.247){\circle*{1}}
\put(101.126, 104.566){\circle*{1}}
\put(101.367, 104.885){\circle*{1}}
\put(101.608, 105.204){\circle*{1}}
\put(101.850, 105.523){\circle*{1}}
\put(102.100, 105.835){\circle*{1}}
\put(102.350, 106.147){\circle*{1}}
\put(102.600, 106.459){\circle*{1}}
\put(102.851, 106.771){\circle*{1}}
\put(103.111, 107.075){\circle*{1}}
\put(103.372, 107.378){\circle*{1}}
\put(103.633, 107.681){\circle*{1}}
\put(103.894, 107.984){\circle*{1}}
\put(104.167, 108.277){\circle*{1}}
\put(104.440, 108.569){\circle*{1}}
\put(104.713, 108.861){\circle*{1}}
\put(104.986, 109.153){\circle*{1}}
\put(105.273, 109.432){\circle*{1}}
\put(105.560, 109.710){\circle*{1}}
\put(105.848, 109.988){\circle*{1}}
\put(106.149, 110.251){\circle*{1}}
\put(106.451, 110.514){\circle*{1}}
\put(106.752, 110.777){\circle*{1}}
\put(107.068, 111.022){\circle*{1}}
\put(107.384, 111.267){\circle*{1}}
\put(107.700, 111.512){\circle*{1}}
\put(108.016, 111.757){\circle*{1}}
\put(108.350, 111.976){\circle*{1}}
\put(108.685, 112.194){\circle*{1}}
\put(109.035, 112.389){\circle*{1}}
\put(109.384, 112.583){\circle*{1}}
\put(109.734, 112.777){\circle*{1}}
\put(110.099, 112.940){\circle*{1}}
\put(110.464, 113.102){\circle*{1}}
\put(110.829, 113.265){\circle*{1}}
\put(111.209, 113.389){\circle*{1}}
\put(111.590, 113.512){\circle*{1}}
\put(111.970, 113.635){\circle*{1}}
\put(112.362, 113.711){\circle*{1}}
\put(112.755, 113.787){\circle*{1}}
\put(113.153, 113.820){\circle*{1}}
\put(113.551, 113.853){\circle*{1}}
\put(113.949, 113.886){\circle*{1}}
\put(114.349, 113.867){\circle*{1}}
\put(114.748, 113.849){\circle*{1}}
\put(115.143, 113.788){\circle*{1}}
\put(115.538, 113.726){\circle*{1}}
\put(115.933, 113.664){\circle*{1}}
\put(116.317, 113.556){\circle*{1}}
\put(116.702, 113.447){\circle*{1}}
\put(117.074, 113.302){\circle*{1}}
\put(117.447, 113.156){\circle*{1}}
\put(117.819, 113.011){\circle*{1}}
\put(118.176, 112.831){\circle*{1}}
\put(118.533, 112.650){\circle*{1}}
\put(118.889, 112.469){\circle*{1}}
\put(119.228, 112.257){\circle*{1}}
\put(119.567, 112.044){\circle*{1}}
\put(119.906, 111.832){\circle*{1}}
\put(120.227, 111.594){\circle*{1}}
\put(120.548, 111.356){\circle*{1}}
\put(120.870, 111.118){\circle*{1}}
\put(121.175, 110.860){\circle*{1}}
\put(121.481, 110.602){\circle*{1}}
\put(121.786, 110.344){\circle*{1}}
\put(122.077, 110.070){\circle*{1}}
\put(122.368, 109.795){\circle*{1}}
\put(122.659, 109.521){\circle*{1}}
\put(122.951, 109.247){\circle*{1}}
\put(123.227, 108.958){\circle*{1}}
\put(123.503, 108.668){\circle*{1}}
\put(123.779, 108.379){\circle*{1}}
\put(124.044, 108.079){\circle*{1}}
\put(124.309, 107.779){\circle*{1}}
\put(124.573, 107.480){\circle*{1}}
\put(124.838, 107.180){\circle*{1}}
\put(125.092, 106.870){\circle*{1}}
\put(125.345, 106.561){\circle*{1}}
\put(125.599, 106.252){\circle*{1}}
\put(125.852, 105.943){\circle*{1}}
\put(126.096, 105.626){\circle*{1}}
\put(126.340, 105.308){\circle*{1}}
\put(126.584, 104.991){\circle*{1}}
\put(126.828, 104.674){\circle*{1}}
\put(127.063, 104.351){\circle*{1}}
\put(127.298, 104.027){\circle*{1}}
\put(127.534, 103.704){\circle*{1}}
\put(127.769, 103.381){\circle*{1}}
\put(127.997, 103.052){\circle*{1}}
\put(128.226, 102.724){\circle*{1}}
\put(128.454, 102.395){\circle*{1}}
\put(128.682, 102.067){\circle*{1}}
\put(128.910, 101.738){\circle*{1}}
\put(129.132, 101.405){\circle*{1}}
\put(129.353, 101.072){\circle*{1}}
\put(129.575, 100.739){\circle*{1}}
\put(129.796, 100.406){\circle*{1}}
\put(130.013, 100.070){\circle*{1}}
\put(130.229, 99.733){\circle*{1}}
\put(130.446, 99.397){\circle*{1}}
\put(130.663, 99.061){\circle*{1}}
\put(130.879, 98.724){\circle*{1}}
\put(131.091, 98.385){\circle*{1}}
\put(131.303, 98.046){\circle*{1}}
\put(131.515, 97.707){\circle*{1}}
\put(131.727, 97.368){\circle*{1}}
\put(131.939, 97.029){\circle*{1}}
\put(132.148, 96.687){\circle*{1}}
\put(132.356, 96.346){\circle*{1}}
\put(132.564, 96.004){\circle*{1}}
\put(132.773, 95.663){\circle*{1}}
\put(132.981, 95.321){\circle*{1}}
\put(133.187, 94.978){\circle*{1}}
\put(133.393, 94.635){\circle*{1}}
\put(133.598, 94.292){\circle*{1}}
\put(133.804, 93.949){\circle*{1}}
\put(134.008, 93.605){\circle*{1}}
\put(134.212, 93.261){\circle*{1}}
\put(134.416, 92.917){\circle*{1}}
\put(134.620, 92.573){\circle*{1}}
\put(134.825, 92.229){\circle*{1}}
\put(135.028, 91.885){\circle*{1}}
\put(135.231, 91.540){\circle*{1}}
\put(135.434, 91.196){\circle*{1}}
\put(135.637, 90.851){\circle*{1}}
\put(135.840, 90.506){\circle*{1}}
\put(136.043, 90.162){\circle*{1}}
\put(136.246, 89.817){\circle*{1}}
\put(136.449, 89.472){\circle*{1}}
\put(136.652, 89.127){\circle*{1}}
\put(136.855, 88.783){\circle*{1}}
\put(137.058, 88.438){\circle*{1}}
\put(137.261, 88.094){\circle*{1}}
\put(137.465, 87.749){\circle*{1}}
\put(137.668, 87.405){\circle*{1}}
\put(137.871, 87.060){\circle*{1}}
\put(138.076, 86.717){\circle*{1}}
\put(138.280, 86.373){\circle*{1}}
\put(138.485, 86.029){\circle*{1}}
\put(138.690, 85.685){\circle*{1}}
\put(138.894, 85.342){\circle*{1}}
\put(139.101, 84.999){\circle*{1}}
\put(139.307, 84.657){\circle*{1}}
\put(139.514, 84.314){\circle*{1}}
\put(139.721, 83.972){\circle*{1}}
\put(139.927, 83.629){\circle*{1}}
\put(140.137, 83.288){\circle*{1}}
\put(140.346, 82.948){\circle*{1}}
\put(140.556, 82.607){\circle*{1}}
\put(140.765, 82.266){\circle*{1}}
\put(140.975, 81.926){\circle*{1}}
\put(141.188, 81.587){\circle*{1}}
\put(141.402, 81.249){\circle*{1}}
\put(141.615, 80.911){\circle*{1}}
\put(141.828, 80.572){\circle*{1}}
\put(142.046, 80.237){\circle*{1}}
\put(142.264, 79.901){\circle*{1}}
\put(142.481, 79.565){\circle*{1}}
\put(142.699, 79.230){\circle*{1}}
\put(142.917, 78.894){\circle*{1}}
\put(143.140, 78.562){\circle*{1}}
\put(143.363, 78.231){\circle*{1}}
\put(143.587, 77.899){\circle*{1}}
\put(143.810, 77.567){\circle*{1}}
\put(144.039, 77.239){\circle*{1}}
\put(144.269, 76.912){\circle*{1}}
\put(144.499, 76.584){\circle*{1}}
\put(144.728, 76.257){\circle*{1}}
\put(144.958, 75.929){\circle*{1}}
\put(145.195, 75.607){\circle*{1}}
\put(145.433, 75.286){\circle*{1}}
\put(145.671, 74.964){\circle*{1}}
\put(145.908, 74.642){\circle*{1}}
\put(146.155, 74.327){\circle*{1}}
\put(146.401, 74.012){\circle*{1}}
\put(146.648, 73.697){\circle*{1}}
\put(146.894, 73.382){\circle*{1}}
\put(147.151, 73.075){\circle*{1}}
\put(147.407, 72.768){\circle*{1}}
\put(147.664, 72.462){\circle*{1}}
\put(147.920, 72.155){\circle*{1}}
\put(148.188, 71.858){\circle*{1}}
\put(148.456, 71.561){\circle*{1}}
\put(148.724, 71.264){\circle*{1}}
\put(148.993, 70.967){\circle*{1}}
\put(149.274, 70.683){\circle*{1}}
\put(149.556, 70.399){\circle*{1}}
\put(149.837, 70.115){\circle*{1}}
\put(150.132, 69.845){\circle*{1}}
\put(150.427, 69.575){\circle*{1}}
\put(150.722, 69.305){\circle*{1}}
\put(151.017, 69.034){\circle*{1}}
\put(151.328, 68.784){\circle*{1}}
\put(151.640, 68.533){\circle*{1}}
\put(151.951, 68.282){\circle*{1}}
\put(152.279, 68.053){\circle*{1}}
\put(152.607, 67.824){\circle*{1}}
\put(152.935, 67.595){\circle*{1}}
\put(153.279, 67.393){\circle*{1}}
\put(153.624, 67.190){\circle*{1}}
\put(153.969, 66.987){\circle*{1}}
\put(154.330, 66.816){\circle*{1}}
\put(154.691, 66.645){\circle*{1}}
\put(155.066, 66.505){\circle*{1}}
\put(155.440, 66.365){\circle*{1}}
\put(155.815, 66.225){\circle*{1}}
\put(156.202, 66.126){\circle*{1}}
\put(156.589, 66.027){\circle*{1}}
\put(156.976, 65.928){\circle*{1}}
\put(157.373, 65.879){\circle*{1}}
\put(157.770, 65.830){\circle*{1}}
\put(158.169, 65.825){\circle*{1}}
\put(158.569, 65.820){\circle*{1}}
\put(158.969, 65.814){\circle*{1}}
\put(159.333, 65.942){\circle*{1}}
\put(159.698, 66.070){\circle*{1}}
\put(159.887, 66.401){\circle*{1}}
\put(160.076, 66.733){\circle*{1}}
\put(160.265, 67.064){\circle*{1}}
\put(160.293, 67.463){\circle*{1}}
\put(160.322, 67.861){\circle*{1}}
\put(160.351, 68.259){\circle*{1}}
\put(160.379, 68.658){\circle*{1}}
\put(160.408, 69.056){\circle*{1}}
\put(160.436, 69.454){\circle*{1}}
\put(160.465, 69.853){\circle*{1}}
\put(160.493, 70.251){\circle*{1}}
\put(160.522, 70.649){\circle*{1}}
\put(160.524, 71.049){\circle*{1}}
\put(160.527, 71.449){\circle*{1}}
\put(160.529, 71.849){\circle*{1}}
\put(160.531, 72.249){\circle*{1}}
\put(160.533, 72.649){\circle*{1}}
\put(160.536, 73.049){\circle*{1}}
\put(160.538, 73.449){\circle*{1}}
\put(160.540, 73.849){\circle*{1}}
\put(160.543, 74.249){\circle*{1}}
\put(160.545, 74.649){\circle*{1}}
\put(160.547, 75.049){\circle*{1}}
\put(160.549, 75.449){\circle*{1}}
\put(160.552, 75.849){\circle*{1}}
\put(160.554, 76.249){\circle*{1}}
\put(160.556, 76.649){\circle*{1}}
\put(160.558, 77.049){\circle*{1}}
\put(160.561, 77.449){\circle*{1}}
\put(160.563, 77.849){\circle*{1}}
\put(160.565, 78.249){\circle*{1}}
\put(160.568, 78.649){\circle*{1}}
\put(160.570, 79.049){\circle*{1}}
\put(160.572, 79.449){\circle*{1}}
\put(160.574, 79.849){\circle*{1}}
\put(160.577, 80.249){\circle*{1}}
\put(160.579, 80.649){\circle*{1}}
\put(160.581, 81.049){\circle*{1}}
\put(160.584, 81.449){\circle*{1}}
\put(160.586, 81.849){\circle*{1}}
\put(160.588, 82.249){\circle*{1}}
\put(160.590, 82.649){\circle*{1}}
\put(160.593, 83.049){\circle*{1}}
\put(160.595, 83.449){\circle*{1}}
\put(160.597, 83.849){\circle*{1}}
\put(160.600, 84.249){\circle*{1}}
\put(160.602, 84.649){\circle*{1}}
\put(160.604, 85.049){\circle*{1}}
\put(160.606, 85.449){\circle*{1}}
\put(160.609, 85.849){\circle*{1}}
\put(160.611, 86.249){\circle*{1}}
\put(160.613, 86.649){\circle*{1}}
\put(160.615, 87.049){\circle*{1}}
\put(160.618, 87.449){\circle*{1}}
\put(160.620, 87.849){\circle*{1}}
\put(160.622, 88.249){\circle*{1}}
\put(160.625, 88.649){\circle*{1}}
\put(160.627, 89.049){\circle*{1}}
\put(160.629, 89.449){\circle*{1}}
\put(160.631, 89.849){\circle*{1}}
\put(160.634, 90.249){\circle*{1}}
\put(160.636, 90.649){\circle*{1}}
\put(160.638, 91.049){\circle*{1}}
\put(160.641, 91.449){\circle*{1}}
\put(160.643, 91.849){\circle*{1}}
\put(160.645, 92.249){\circle*{1}}
\put(160.647, 92.649){\circle*{1}}
\put(160.650, 93.049){\circle*{1}}
\put(160.652, 93.449){\circle*{1}}
\put(160.654, 93.849){\circle*{1}}
\put(160.657, 94.249){\circle*{1}}
\put(160.659, 94.649){\circle*{1}}
\put(160.661, 95.049){\circle*{1}}
\put(160.663, 95.449){\circle*{1}}
\put(160.666, 95.849){\circle*{1}}
\put(160.668, 96.249){\circle*{1}}
\put(160.670, 96.649){\circle*{1}}
\put(160.672, 97.049){\circle*{1}}
\put(160.675, 97.449){\circle*{1}}
\put(160.677, 97.849){\circle*{1}}
\put(160.679, 98.249){\circle*{1}}
\put(160.682, 98.649){\circle*{1}}
\put(160.684, 99.049){\circle*{1}}
\put(160.686, 99.449){\circle*{1}}
\put(160.688, 99.849){\circle*{1}}

\end{picture}
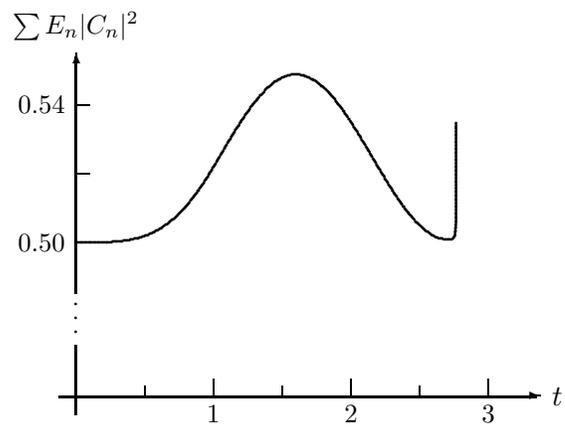
\end{document}